\documentclass{PoS}

\newcommand{\veritas}{\textit{VERITAS}}
\newcommand{\magic }{\textit{MAGIC}}

\newcommand{\onees}{\textit{1ES~0806+524}}

\title{\veritas\ long-term (2006-2014) observations of the BL Lac object \onees}

\ShortTitle{\veritas\ long-term (2006-2014) observations of the BL Lac object \onees}

\author{\speaker{Matteo Cerruti}\ for the \veritas\ Collaboration\\
        Harvard-Smithsonian Center for Astrophysics, 60 Garden St., Cambridge, MA 02138, USA\\
        E-mail: \email{matteo.cerruti@cfa.harvard.edu}}

\abstract{The discovery of the high-frequency-peaked BL Lac object 1ES 0806+524 (z=0.138) as a source of very-high-energy (VHE, E>100 GeV) gamma-ray photons was announced in 2008 by the \veritas\ Collaboration, at a level of 1.8\% of the Crab Nebula flux above 300 GeV. Since then, \veritas\ has continued observing the source over multiple seasons, significantly improving the significance of the detection. We report the results of the analysis of the 2006-2014 \veritas\ data, corresponding to a total exposure of about 70 hours. We present the new, average VHE spectrum of the source, together with the multi-year light-curve constraining long-term VHE variability.}

\FullConference{The 34th International Cosmic Ray Conference,\\
		30 July- 6 August, 2015\\
		The Hague, The Netherlands}

\begin{document}

\section{Introduction}

Blazars are the most common source of $\gamma$-ray photons in the extra-galactic sky. The current generation of imaging atmospheric Cherenkov telescopes (IACTs), observing the universe at very-high-energies (E$>100$ GeV; VHE), has identified around 60 blazars as VHE emitters\footnote{See tevcat.uchicago.edu for an updated list of VHE sources}.  Among the different blazar subclasses (BL Lac objects and flat-spectrum radio-quasars, FSRQ; further divided into low/intermediate/high-synchrotron peaked blazars), the VHE extra-galactic sky is largely dominated by high-synchrotron-peaked BL Lac objects. A peculiarity of VHE astronomy is that the universe is opaque to VHE photons, due to pair-production triggered by the interaction of $\gamma$-rays with the extra-galactic background light (EBL) in visible and infrared light. The effect is that VHE astronomy is limited in redshift,  the current most distant VHE source being the gravitationally-lensed blazar \textit{S3~0218+35}, at z=0.94. This implies that the majority of VHE blazars are nearby high-synchrotron-peaked BL Lac objects.\\ The variability properties of VHE blazars are still poorly understood, some blazars being stable sources of VHE photons, while others having been detected only during rapid flaring episodes. The \veritas\ collaboration is carrying on a long-term monitoring of several VHE blazars \cite{Dumm13}, with the goal of studying the long-term evolution of the VHE flux, possibly associated with spectral variability. \\

The BL Lac object \onees\ was one of the first discoveries of the current generation of IACTs, detected by \veritas\ during observations in 2006-2008 \cite{0806Veritas}. 
In this work, we present the results of \veritas\ observations of \onees\ from 2006 to 2014, re-analyzing as well part of the data published in \cite{0806Veritas}.\\
During February 2011, the \magic\ collaboration has reported the detection of a rapid (day-long) flare from \onees, as discussed in \cite{0806Magic}. The \veritas\ telescopes observed the source before and after the \magic\ flare, and we present as well the detailed results of the \veritas\ observations during 2011.\\

\section{\veritas\ data analysis}

\veritas\ (Very Energetic Radiation Imaging Telescope Array System) is an array of four 12-meter diameter telescopes, located at the base of Mount Hopkins, in Southern Arizona, at the Fred Lawrence Whipple Observatory. The array is arranged in a diamond configuration, with sides of about 100m. Each telescope is equipped with a camera made of 499 photo-multiplier tubes, sensitive to the weak Cherenkov light emitted by the cascades produced in the interaction of VHE photons with the Earth's atmosphere. \\
The system is able to reconstruct $\gamma$-ray emission from about 85 GeV to $>$30 TeV, with an energy resolution of 15-25\%. The field of view of the instrument is $3.5^\circ$, with an angular resolution $R_{68\%} < 0.1^{\circ}$ at 1 TeV. The system is able to detect a source with a flux level comparable to the Crab nebula in 12 minutes, and a source with a flux level of the order of 5\% of the Crab nebula in 1.5 hours.\\
The \veritas\ array, in operation since 2006, has gone through several upgrades during its scientific operations: in 2009 one of the four telescopes was relocated in order to get a symmetric array; in 2012 the PMTs in the camera have been replaced, achieving a better sensitivity and a lower energy threshold. For further details see \cite{Holder, Kieda}.\\

The observations of \onees\ presented in this proceedings include data taken before/after the relocation of one of the telescopes, and before/after the camera upgrade. No observations taken under bright-moonlight conditions \cite{Sean,1727} are included in the data-set.
The data-set comprises 72 hours of \veritas\ exposure (after taking into account the dead-time of the instrument). Compared to the results presented in \cite{0806Veritas}, we excluded from the data-set the observations taken in 2006 using only two of the four \veritas\ telescopes.\\

All observations were taken using the \textit{wobble} observing mode, where the telescopes are pointed 0.5$^\circ$ away from the target, to allow a simultaneous estimation of the emission from the source and the background. The discrimination between Cherenkov showers triggered by $\gamma$-rays or cosmic rays is done using standard cuts on the images, as described in \cite{Holder}.\\

For the data-set as just described, the VHE emission from \onees\ is detected with the \veritas\ instrument at a level of 7.1 standard deviations over the background, as estimated following \cite{LiMa}, above an energy threshold of 300 GeV.\\

\subsection{Spectral reconstruction}

\begin{figure*}
	   \centering
		\includegraphics[width=250pt]{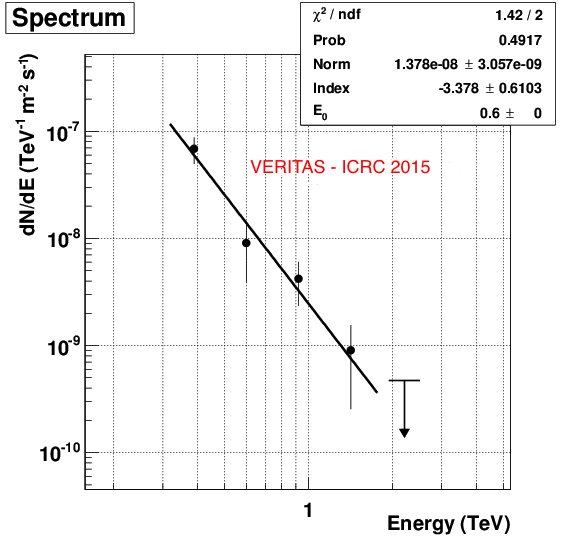}
	  \caption{Spectrum of \onees\ as measured by \veritas.\label{spectrum}}
   \end{figure*}

The previously published VHE spectrum of \onees, as presented in \cite{0806Veritas}, was parametrized by a power-law function between 300 and 700 GeV:
\begin{equation}
\Phi(E) [\textrm{cm}^{-2}\ \textrm{s}^{-1}\ \textrm{TeV}^{-1}] = K \left(\frac{E}{400\ \textrm{GeV}}\right)^{-\Gamma} 
\end{equation}
with $K=(6.8 \pm 1.7)\times 10^{-12}\  \textrm{cm}^{-2}\ \textrm{s}^{-1}\ \textrm{TeV}^{-1}$ and $\Gamma=3.6 \pm 1.0$.\\
Using the 2006-2014 data-set, the new average \veritas\ spectrum of \onees\ extends from 300 GeV to  1.8 TeV, and can be parametrized by a power-law function
\begin{equation}
\Phi(E) [\textrm{cm}^{-2}\ \textrm{s}^{-1}\ \textrm{TeV}^{-1}] = K \left(\frac{E}{600\ \textrm{GeV}}\right)^{-\Gamma} 
\end{equation}
with $K=(1.4 \pm 0.3)\times 10^{-12}\  \textrm{cm}^{-2}\ \textrm{s}^{-1}\ \textrm{TeV}^{-1}$ and $\Gamma=3.4 \pm 0.6$.
The new result is consistent with the previous \veritas\ result, but extends to higher energies, and has smaller uncertainties. The errors on the spectral reconstruction are statistical only. In Figure \ref{spectrum} we show the average spectrum of \onees\ as measured by \veritas.\\

                   \begin{figure*}
	   \centering
		\includegraphics[width=240pt]{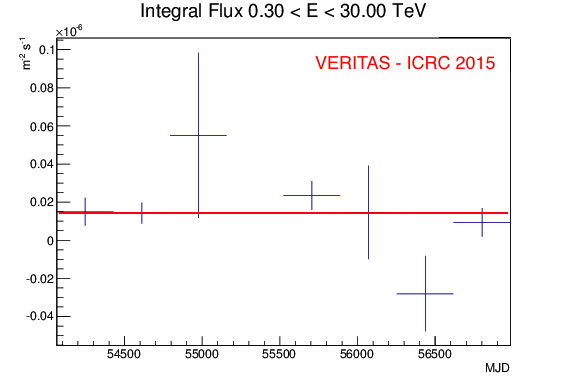}
		\includegraphics[width=240pt]{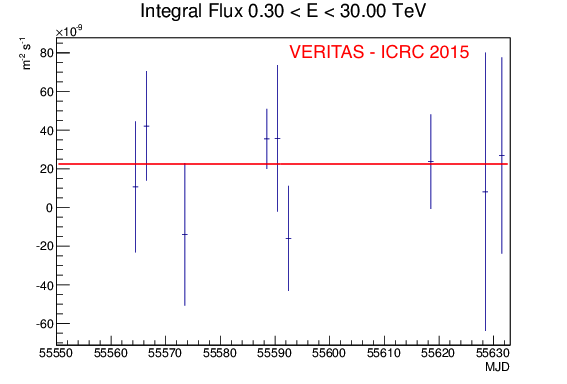}
	  \caption{Light-curve of \onees\ as measured by \veritas. \textit{Top:} year-by-year light-curve, from 2006 to 2014; \textit{Bottom:} day-by-day light-curve, zoomed on the 2011 season. \label{LC}}
   \end{figure*}

\subsection{Temporal variability}

In Figure \ref{LC} we show the VHE integral light-curve of \onees, estimated above the \veritas\ energy threshold (300 GeV). In the top panel we present the long-term (2006-2014) light-curve rebinned per observing season (September to June). No evidence of variability is seen in the data: the flux from \onees\ is compatible with constant emission ($\chi^2= 7/6$), and the average integral flux is $(1.4 \pm 0.3)\times 10^{-12}\  \textrm{cm}^{-2}\ \textrm{s}^{-1}$, corresponding to 1.2$\% \pm 0.3\%$ of the flux from the Crab nebula \cite{Hillas98}.\\

During February 2011, \onees\ showed rapid VHE flaring activity, as reported in \cite{0806Magic}. The \veritas\ telescopes observed the source before and after the \magic\ flare, showing a flux in agreement with the average VHE emission. The day-by-day \veritas\ light-curve of \onees\ during 2011 is shown in the bottom panel of Fig. \ref{LC}. This result supports the fact that the $\gamma$-ray flare observed by \magic\ was an isolated event, happening on a short time-scale. 

\section{Conclusions}

In this work we reported the results of long-term (2006-2014) observations of the BL Lac object \onees\ using the \veritas\ telescopes. The flux and spectrum of the source is consistent with the previous \veritas\ publication, which used data from 2006 to 2008, but the higher statistics allowed a better spectral reconstruction, extending to higher energies.
The long-term light-curve is consistent with a constant flux, at a level of about $1.2\%$ of the Crab nebula flux above 300 GeV. We also investigated in detail the daily light-curve during the 2011 observing season, before and after the VHE flare observed by the \magic\ telescopes, without detecting any evidence of enhanced emission. This result supports the fact that the February 2011 flare seen by \magic\ was characterized by a short time-scale, on the order of a day.\\

\end{document}